# Turbulence on the Global Economy influenced by Artificial Intelligence and Foreign Policy Inefficiencies

*Kwadwo Osei Bonsu*
*(corresponding author)*
*k.oseibonsu@pop.zjgsu.edu.cn,*
*School of Economics,*
*School of Law and Intellectual Property*
*Zhejiang Gongshang University*

*Professor Jie Song*
*School of Law and Intellectual Property*
*Zhejiang Gongshang University*

**Abstract:**

It is said that Data and Information are the new oil. One, who handles the data, handles the emerging future of the global economy. Complex algorithms and intelligence-based filter programs are utilized to manage, store, handle and maneuver vast amounts of data for the fulfilment of specific purposes. This paper seeks to find the bridge between artificial intelligence and its impact on the international policy implementation in the light of geopolitical influence, global economy and the future of labor markets. We hypothesize that the distortion in the labor markets caused by artificial intelligence can be mitigated by a collaborative international foreign policy on the deployment of AI in the industrial circles. We, in this paper, then proceed to propose a disposition for the essentials of AI-based foreign policy and implementation, while asking questions such as 'could AI become the real Invisible Hand discussed by economists?'.

## Introduction

Artificial Intelligence is also known as Machine intelligence as it is based on complex computer algorithms. It can be defined as any mechanism or phenomenon that perceives its environment and takes actions that maximize its chance of successfully achieving its goals[1]. The stimulus of information is called Data.

The role of the internet is growing exponentially in both governmental and non-governmental organizations. Artificial Intelligence has a wide range of applicability in the Governmental ministries. Machine Learning which is basically a branch of AI, along with natural language processing and automated decision making with the help of stimulated data could substitute the human man power. It can handle all the tasks that require human mind aptitude to perform. Artificial intelligence can be perceived as non-biological form of intelligence in this era; there could be biologically induced AI in the future with advancement in quantum computing and gene editing technology. Artificial Intelligence has contributed in every sphere of human life right from the home appliances to transportation by introducing self-driving cars. The application of robotics on the assembly line of many industrial complexes has also been a plus for mankind in expediting manufacturing processes. The arena of Military strategies and defense forces are also enshrined with modern compact artificial intelligence-based drones that fly, combat and refill without any human effort, although there are international regulations preventing the use of autonomous weapons. Currently many military organizations along with extremist groups are using AI-based drones[2]. This calls for tighter international regulation on use of automatic weapons and the use of drones as a whole. Artificial intelligence will open the frontiers of human understanding at both macro and micro levels. It will enhance our existing rationality and help us cope with all the existing problems more effectively. Artificial intelligence also reshapes all our economic, political and social preferences, it is a powerful tool that can manipulate human behavior in most unpredicted ways. Though the development of artificial intelligence is at infancy, we cannot ignore the potentiality of threat it may pose financially, politically, economically and culturally (the extent of potential threat isn't the focus of this paper). Governments must regulate and follow up on research and development made in AI and respond in the best possible way to ensure maximization of utility to minimize disutility.

Artificial Intelligence is widely used in both external and internal government affairs such as improved public management, urban planning, banking, security, transportation and so on. The design, building, use, and evaluation of algorithms and computational techniques is improving at an alarming rate[3]. Data is the major entity of the era of artificial intelligence. The more data governments and public institutions manage to integrate into their systems, the higher the capabilities of machine learning to make decisions based on this data will be[4]. The authenticity of data is directly proportional to the outcome of the process. Data manipulations may cause inverse effects to the potential outcomes.

The paper intends to provide the detailed effects of artificial intelligence over the government and on the international policy implementation. Every implementation action can influence policy problems, resources, and objectives as the process evolves.[5]

**Artificial Intelligence and Governments**

The development of Artificial intelligence can be traced back into the 20th century. In the Second World War the U.S Government was actively using artificial intelligence-based programs in marine aircrafts[6] though they were not as sophisticated as the modern appliances of artificial intelligence.

Recent studies show that more and more human behavior can be predicated by artificial intelligence like the Google Maps suggestions of the fastest route based on data of personal Smartphone is a notable example of artificial intelligence in our daily lives. Moreover, applications of AI in the government have been created for crime risk. Kouziokas confers crime risk prediction as a factor that contributes to safer travel in urban areas using artificial neural networks - an example of AI[7]. The data will calculate high crime rate areas using local police data bases the data was organized into cluster and precisely determine the risk factor of crime in a certain area and time of a metro pole these findings would assist to frame the policies to counter the crime and improving the governability of the state. Many of such type of initiatives would be introduced by Government in all the public sectors to curtail the lag of decision delay and commitment and enhanced the efficiency of the government. This public policy approach regulates the extent by which the applications of artificial intelligence have impact on public lives. Birckland considers the public policy-cycle as an appropriate framework for understanding the complexity of the public decision-making processes, as well as the actors involved in such processes. Thus, activities, actors and drivers of public policy can be considered within this framework[8].

In terms of Foreign policy, Artificial intelligence has opened new vistas of science and technology but many of the existing traditional diplomacy can be adapted into this new field while planning for the significant changes in the coming time. The potential of Artificial intelligence to bring change in this world has initiated a competition among different governments of the states to gain the strategic advantage. China's national AI strategy shows how seriously governments take this technology China is placing major bets on the development of Artificial intelligence and Robotics. In a recent speech, Russian President Vladimir Putin stated bluntly that the country that gains an edge in AI will be the ruler of the world[9] that shows the heated races between the national states to gain the maximum stake at the artificial intelligence.

There are chances and preventive measures that ought to be taken before the artificial intelligence fully indulge in human lives. The steps taken are discussed under the two different approaches one of them is aggressive in nature whereas the other steps can be taken at pragmatic level. The initial development requires massive reformation in the artificial intelligence these are more decisive steps and determine the scope of the foreign policies. It would ultimately mean that we would witness the major change in human work as diplomat. These changes will take a long time to observe hence they are merely the spectrum of our discussion; we are more confined about the pragmatic view of artificial intelligence. All the frontiers of the ministries undergo a major shift because the segments of the artificial intelligence revolutionizes all the existing subjects that Government needs an immediate policy for them. Governments should primarily focus on assets and resources while keeping in mind the more significant planning of the near future events. Many of the existing techniques of diplomacy can be adapted to the rising artificial intelligence. While the current techniques of artificial intelligence can assist us to initiate, this pragmatic approach of artificial intelligence it does not preclude thinking about decisive changes that the technological invention might require for our foreign policy institutions. The method of policy making ultimately choose to govern the wide range of Artificial Intelligence applications and it will have a decisive effect on the final set of opportunities and benefits that could resultantly come from artificial intelligence.

This paper does not intend to counter all the claims that may arise after the advent of artificial intelligence but limits its scope to matters concerned with artificial intelligence and public policy and the implementation of international policy.

The primary purpose of this paper is to suggest an outline of the solutions of all the emerging complexities in the foreign policy implementation with the advent of artificial intelligence.

**AI Policy Propositions and Regulations**

The "lenient" policies from the Government to regulate the usage and the development of artificial intelligence would be key a factor for the growth of artificial intelligence. The innovation policies would be permission less and swift if we want to capture a massive breakthrough of benefits of the artificial intelligence and robotics. In the case of internet, the world had adopted the most lenient policy making view against has helped to spur the digital revolution; government legislators can extend that ethos to new sectors of artificial intelligence to attain similar technological outcomes[10].   The challenges facing the development of artificial intelligence are so diverse that their solutions alarm the public concern.

    Firstly, the government and policy makers must distinguish the existing segments of the artificial intelligence on the basis of their utility and development techniques so that they will decide and segregate each segment of artificial intelligence development and enforce the appropriate steps successfully. For example, the regulations and preventive steps for artificial intelligence that intends to address experimental medical applications, should not inadvertently be applied to social media applications of artificial intelligence because of broad or improper wording and lack of definite segregation between developmental projects. On the other hand some developmental programs of artificial intelligence could be exempted from all the policy making barriers and regulations due to their utility and sensitivity and the programs of artificial

intelligence that pose direct risks to the safety of the society or health of the public could be examined to conclude what kind of oversight regulations are appropriate to minimize their potential risk in artificial intelligence.

    Secondly, Governments and policy makers should figure out all the potential threats and concerns in a productive form to avoid the worst-case scenarios while deciding about the fate of the technological advancement of artificial intelligence. Governments and policy makers have an authority to indulge an excessive approach while engineering the rules and regulations for artificial intelligence. This precautionary principle of Governments to manage the 'threat' pronounced against artificial intelligence chronicles for-better-or-for-worse outcomes for the humans who invoke them.[11] Governments should take care to separate fantasy from reality when addressing automated technologies.

    Finally, governments and policy makers should adopt a lenient view against the artificial intelligence technologies while legislating for the future regulations of these technologies so the world will benefit more from the artificial intelligence. The lesson of the contemporary policy approaches to the internet in the 1990s is sufficient guidebook for the regulations of artificial intelligence. Policymakers in both of the United States and European union follow finical models. The examples of the United States prioritize a clear space for experimentation and commercialization begets collaboration and growth of the internet and industry. Whereas the European union appears to be inadvertently quashing the industry before it has the chance to develop making it dull in terms of utility.

The state of patience and general openness and lenient policies in the development and innovation of the technologies are the wise decisions because they provide ample breathing space for the nourishment of future technologies of artificial intelligence and spare a time frame for the governmental organizations to monitor and see the public response and attitude of the societies as the artificial intelligence evolves over the time.

**Foundations of AI based Foreign Policy**

The policy response of the foreign ministries to respond over the growing number of changes is concluded into four broad structural outcomes that we draw from the assessment discussed below:

We have to adopt a swifter approach in terms of problem solving and work ambitiously to integrate the technological knowledge with our existing conventional work load fashion. It would require substantial reorganization of the institutes. The pragmatic strategy is to claim the limits of the work and benchmark that was achieved in the digital era of the internet. For Example: The small offices that are enough to operate a cyber network are not sufficient to deliver the same mark of effective management and development when it comes to artificial intelligence. All the operations in the foreign offices that make through the artificial intelligence that may be centrally coordinated but they required special set of technicians to work and operate. It is concluded that if we prioritize the development of the artificial intelligence and new technologies, we must shift the focus and de-prioritize all the other existential issues with decreasing relevance.

The second lesson for effective enforcement of artificial intelligence in foreign policy agenda is the swift and effective response for the network actor of the private companies, research institutes and civil society. For that purpose, an effective knowledge of data is required to identify the most useful interventions and to overcome the effect of double occurrence by establishing a series of collaborations we draw in our existing figures. This can be illustrated from the example which artificial intelligence policy issues are corresponding in terms of the arms control efforts of the cold war and the period prior to cold war[12].

The third lesson motivate us to adopt a problem-solving attitude while handling our foreign policy response; this actually means to avoid the bureaucratic glitches in handling specific issues. The development of artificial intelligence is faster beyond the response rate of existing bureaucratic framework of the world as it requires conscious effort in adopting to work structures. The existing problem-solving process is axiomatic. The mighty software industry that is responsible for the handling of the development of the technology of artificial intelligence and robotics and many of its features can be usefully applied in policy development.

The last lesson is that Humans should be ready for the persisting challenges and difficulties which may arise during the process; these challenges can be seen to be mainly more consent with the departments of H.R, Human Resource management sectors. The man power skilled to work with artificial intelligence is required to meet the needs of our diplomatic practices. The conventional process of hiring and recruiting suitable candidates for the jobs in ministries of foreign affairs needs to be in tandem with the standard operations and procedure of Governments; young technology experts should be hired and involved in the crucial affairs of ministries. The challenge of finding, selecting and recruiting the artificial intelligent experts who are at the same time skilled in social science and law and then integrating them into the required posts in the required offices and embassies is not easy task. The best and the most suitable candidates for these posts may not come through the conventional mode of foreign and civil service officers and it may require a separate aptitude test of different qualities for their enrollment.

**Essentials of AI based foreign policy and The Global Economy**

The advancement of artificial intelligence has also brought about some managerial as well as social challenges international relations. The essentials of foreign policy required for harmonious synchronization of artificial intelligence into foreign policy rules are discussed below.

The first and foremost priority of the foreign ministries and offices is to calculate the magnitude of the major issues that would be directly caused by the development of artificial intelligence. We cannot shape the future dilemma of artificial intelligence without setting the preferences of our research. The stakes of this technology are well understood by the nations of the states as it can be concluded from the fact that the recently published New Generation of Artificial Intelligence Development Plan, China expects to generate $59 billion in artificial intelligence-based technology[13]. The Government of South Korea initiated an investment of $863 million in artificial-intelligence technology[14] and the many governments have allocated huge funds for the development and research of AI[15]. This paper asserts that the initiative by the world governments is fueling the unending development of the technology under discussion and shows that policy making plays a vital role in shaping the advancement of technology in AI.

However, too much attention on AI would suffocate many other areas of research which is detrimental to the development of Agriculture, Materials Science, Biology, Pharmacy etc.

The art of diplomacy is a tool of effective communication. The media and civil societies are a source of public narrative in a society. The awareness regarding the advancement of artificial intelligence should be the part of a new communication. Artificial Intelligence has seen an acceleration of advances in the field of medicine, health, economics, finance, security and energy[16]. These advancements were possible through Machine Learning and Data analysis obtained through the daily activities of societies (there are privacy issues which will not be discussed in this paper).

The agreements are mutually consensus hence the world must engage in a productive dialogue about the future of the development of artificial intelligence. The process of dialogue will suggest new strategic challenges that would put states and international organizations onto the same page; thereby easing the strain in international relations according caused by the transfer of technology. This model is very effective in the development of regulatory measures in governing the internet, cyber security and the digital economy.

The path of development of artificial intelligence begins with the new standards of confidence building between the international policies in the groups of multilateral experts[17]. These series of meetings would be the foundation of international law that governs the applications of artificial intelligence. Many of the short term and long-term agreements would be signed to minimize the impact of artificial intelligence into the labor market because the introduction of AI and highly advanced robots will render a huge number of workers jobless while introducing new jobs.[18] This paper disagrees with Sachs et al that both the human capital and labor workers would devalue and that only highly skilled artificial intelligence experts would retrieve their value[19]. Government regulation is required to make sure a gradual integration of AI is executed in order to reduce the amount of turbulence caused by AI in distorting the label markets.

Foreign offices pose as the leading figures by collaborating between all the stakeholders from different regions and sectors to sort out all the challenges and build a framework to avail all the opportunities of artificial intelligence. It would be recalled that the initial conferences titled on the freedom and free use of internet resulted in strengthening the scope of the right based policy agenda. Similar agenda to promote consumer rights should be initiated. This work can be based on the footprints of public diplomacy.

The introduction of a competent global body should be prioritized in advocating the narrative of global civil societies regarding the implication of the development of artificial intelligence and international policy making. The majority of the general public has had the stuff of science fiction movies in their minds regarding the implications of artificial intelligence featuring robotic killings and algorithms empowered to deliver lethal forces. In Argentina on the date of July 28, 2015, an open letter was written to the Joint conference on the development of robotics and artificial intelligence for imposing a ban on autonomous weapons[20]; that letter was signed by more than 3000 scientist and artificial intelligent researchers[21]. These narrative platforms also address public grievances and assist foreign bodies in policy implementation.

The foreign offices and all the embassies should assist in integrating all the available information and form a network to collaborate and share information between the different segments of their offices. All the major breakthroughs in the development of artificial intelligence must be reported on priority basis and all the developmental programs of artificial intelligence must be monitored. The head of the missions and project directors should be engaged with their delegations. All the opportunities and threats must be equally inspected and shared. Facial recognition and image processing where data is trained based on algorithms on the principles of information gathering has been a race in China among both top and upcoming tech giants such as Baidu, Tencent, and Sensetime etc. [22] [23]. This is a two-edged sword. However, information sharing helps to foster cooperation thereby optimizing economic development.

**Methods**

This paper uses qualitative conceptual analysis in diving into the issues surrounding artificial intelligence and foreign policy. We did cross referencing of documentation regarding both AI, international law, international economic law, foreign policy, globalization, machine learning and sought into both the anthological and contemporary phenomena emerging in the relationship among AI, foreign policy and labor markets.

**Results**

It can be inferred from our research that the collusion of artificial intelligence and the implementation of foreign policy is quite under developed. Much research is required in artificial intelligence to develop applications usable in foreign policy implementation, whereas much research is needed in the area of international relations and international law in showing how the inevitable advancement of AI can be regulated without impeding on its progress. The significance of diplomacy and statecraft in terms of artificial intelligence is scare because there is the lack of sufficient number of AI experts with interests and opportunities taking key positions in foreign ministries of governments. Also, there are fewer international relations and law experts with in depth training in computer science and AI. Disparateness between technology and the social sciences causes discomfiture in implementing foreign policies relating to AI.

**Discussion**

This paper has discussed a brief analysis and presented guidelines on making diplomatic practices relating to artificial intelligence. As we have seen, there are institutes operating under constraints whether being political, bureaucratic, or financial constraints, this paper has given a pragmatic view of the implementation of foreign policy on artificial intelligence such as the utilization of the existing tools of diplomacy featuring a fraternity of systematic adoptions. With a huge chunk of the global economy moving towards digitization, "Cyber foreign policy" should be given a prime attention and discussed among both governmental and non-governmental organizations especially in the area of AI. Consumer right protection should be imperative to policy makers.

**Conclusion**

As shown, advancement in AI and robotics is frustrating the usability of a portion of the labor force as well as not replacing but creating news jobs. The turbulence of this phenomenon causes distortion in the labor markets thereby shifting the supply and demand curves of labor. The global economy, geopolitics and international relations are affected dramatically as we can see; the trade war between China and The USA is partly based on the shifting of labor from the later to the former, due to the fact that most jobs in the USA, the EU and Japan have shifted from intensive labor to service based and technology based which has boosted the manufacturing sector in China. Advancement in AI has caused many factories to replace repetitive jobs in the production line with automation. Most of these factory workers have moved onto new jobs created by Uber, Didi, Meituan and many others. While others are video streaming and selling products on Tiktok. These are examples that show that new jobs are created whiles others are being lost.

During the industrial revolution, horse groomers probably were worried about loosing their jobs while new jobs like mechanics were being created. It is a 'hand-go-hand-come' tale. Artificial intelligence could create a new political revolution where everything is governed by itself through a real 'Invisible Hand'. It could (if not become) perform the actions of the invisible hand which economists have theorized for decades. Governments and educational institutions should come up with programs that train a breed of technology, AI, law, international relations and economics hybrids of experts to be given key policymaking roles in multi-national endeavors.

**Abbreviations**

AI: Artificial Intelligence

U.S: United States

EU: European Union

**Figure legends**: Not applicable

**Figures**: Not applicable